\documentclass[%
 reprint,
superscriptaddress,
 amsmath,amssymb,
 aps,
prl,
]{revtex4-2}

\usepackage{graphicx}
\usepackage{dcolumn}
\usepackage{bm}

\usepackage[colorlinks=true, allcolors=blue]{hyperref}

\begin{document}

\preprint{APS/123-QED}

\title{Magnetoresistance Oscillations in Few-Layer NbSe\textsubscript{2} in Superconducting Fluctuation Regime}

\affiliation{State Key Laboratory of Quantum Functional Materials, Department of Physics, and Guangdong Basic Research Center of Excellence for Quantum Science, Southern University of Science and Technology, Shenzhen 518055, China}
\affiliation{Quantum Science Center of Guangdong-Hong Kong-Macao Greater Bay Area, Shenzhen 518045, China}

\author{Xiaolong Yin}
\thanks{These authors contributed equally to this work.}
\affiliation{State Key Laboratory of Quantum Functional Materials, Department of Physics, and Guangdong Basic Research Center of Excellence for Quantum Science, Southern University of Science and Technology, Shenzhen 518055, China}

\author{Congzhe Cao}
\thanks{These authors contributed equally to this work.}
\affiliation{State Key Laboratory of Quantum Functional Materials, Department of Physics, and Guangdong Basic Research Center of Excellence for Quantum Science, Southern University of Science and Technology, Shenzhen 518055, China}

\author{Yibin Feng}
\affiliation{State Key Laboratory of Quantum Functional Materials, Department of Physics, and Guangdong Basic Research Center of Excellence for Quantum Science, Southern University of Science and Technology, Shenzhen 518055, China}

\author{Kenji Watanabe}
\affiliation{Research Center for Functional Materials, National Institute for Materials Science, 1-1 Namiki, Tsukuba 305-0044, Japan}

\author{Takashi Taniguchi}
\affiliation{International Center for Materials Nanoarchitectonics, National Institute for Materials Science, 1-1 Namiki, Tsukuba 305-0044, Japan}

\author{Jiawei Mei}
\thanks{Corresponding author: meijw@sustech.edu.cn}
\affiliation{State Key Laboratory of Quantum Functional Materials, Department of Physics, and Guangdong Basic Research Center of Excellence for Quantum Science, Southern University of Science and Technology, Shenzhen 518055, China}

\author{Qi-Kun Xue}
\thanks{Corresponding author: xueqk@sustech.edu.cn}
\affiliation{State Key Laboratory of Quantum Functional Materials, Department of Physics, and Guangdong Basic Research Center of Excellence for Quantum Science, Southern University of Science and Technology, Shenzhen 518055, China}
\affiliation{Quantum Science Center of Guangdong-Hong Kong-Macao Greater Bay Area, Shenzhen 518045, China}
\affiliation{Department of Physics, Tsinghua University, Beijing 100084, China}

\author{Shuo-Ying Yang}
\thanks{Corresponding author: yangsy@sustech.edu.cn}
\affiliation{State Key Laboratory of Quantum Functional Materials, Department of Physics, and Guangdong Basic Research Center of Excellence for Quantum Science, Southern University of Science and Technology, Shenzhen 518055, China}
\affiliation{Quantum Science Center of Guangdong-Hong Kong-Macao Greater Bay Area, Shenzhen 518045, China}

\date{\today}

\begin{abstract}
Quantum interference phenomena in superconductors, such as Josephson interference and Little-Parks oscillations, serve as powerful probes of phase coherence, symmetry breaking and vortex dynamics. However, they are typically observed in well-defined mesoscopic structures, and their behavior in the two-dimensional limit remains largely unexplored. Here, we report periodic magnetoresistance oscillations, superconducting interference patterns, and interfering diode effect in unpatterned few-layer NbSe$_2$. These phenomena emerge exclusively within the superconducting fluctuation regime of thin samples, consistent with the enhanced anomalous metallic behavior of atomically thin NbSe$_2$. The non-monotonic temperature dependence of both the oscillation amplitude and the diode efficiency can be captured by a model in which thermally activated vortices traverse intrinsic supercurrent loops. Our results reveal that the observed interference phenomena originate from the lost of global phase coherence, providing a new route to accessing interference effects in unpatterned superconductors.
\end{abstract}

\maketitle



Magnetic flux quantization is one of the most fundamental properties of superconductors and represents a macroscopic consequence of Cooper pair condensation. The pioneering experiment of Little and Parks demonstrated that the superconducting transition temperature $T_{c}$ oscillates with applied magnetic field in a superconducting cylinder, reflecting the periodic variation of the superconducting free energy\cite{PhysRevLett.9.9}. In mesoscopic superconducting rings, this modulation is directly revealed as magnetoresistance oscillations near the superconducting transition temperature. More recently, periodic magnetoresistance oscillations and superconducting interference patterns have also been reported in several intrinsic superconductors without nanopatterned structures. These observations have been interpreted as signatures of emergent Little–Parks–like effects arising from intrinsic supercurrent loops, potentially associated with mechanisms such as chiral superconducting domains in CsV\textsubscript{3}Sb\textsubscript{5}\cite{Le2024,yqth-sfm8,lou2025radio,blom2025emergentnetworkjosephsonjunctions}and KV\textsubscript{3}Sb\textsubscript{5}\cite{wu2025supercurrent}, chiral edge modes in MoTe\textsubscript{2}\cite{doi:10.1126/science.aaw9270}, charge ordering in ionic-gated TiSe\textsubscript{2}\cite{Li2016,Liao2021} and underdoped Bi\textsubscript{2}Sr\textsubscript{2}CaCu\textsubscript{2}O\textsubscript{8+x}\cite{Liao2022}.

In recent years, 2D superconductivity has emerged as a fertile platform for discovering quantum phenomena that are inaccessible in bulk materials. In the 2D limit, enhanced phase fluctuations and reduced dimensionality fundamentally reshape the coherence and collective behavior in atomically thin materials. Vortex dynamics lies at the core of what distinguishes 2D superconductors from their three-dimensional counterparts. The Berezinskii–Kosterlitz–Thouless (BKT) transition is governed by the unbinding of vortex–antivortex pairs\cite{berezinskii1972destruction,kosterlitz1973ordering,halperin1979resistive,PhysRevLett.42.1165}, while the weakened screening in ultrathin superconductors allows magnetic field to penetrate over the Pearl length, producing vortices that are more expanded than in 3D superconductors\cite{pearl1964current}. Nevertheless, the interplay among dimensionality, phase coherence, and vortex dynamics, particularly their impact on superconducting interference phenomena, remains poorly understood.
\begin{figure*}[t]
    \centering
    \includegraphics[width=0.8\textwidth]{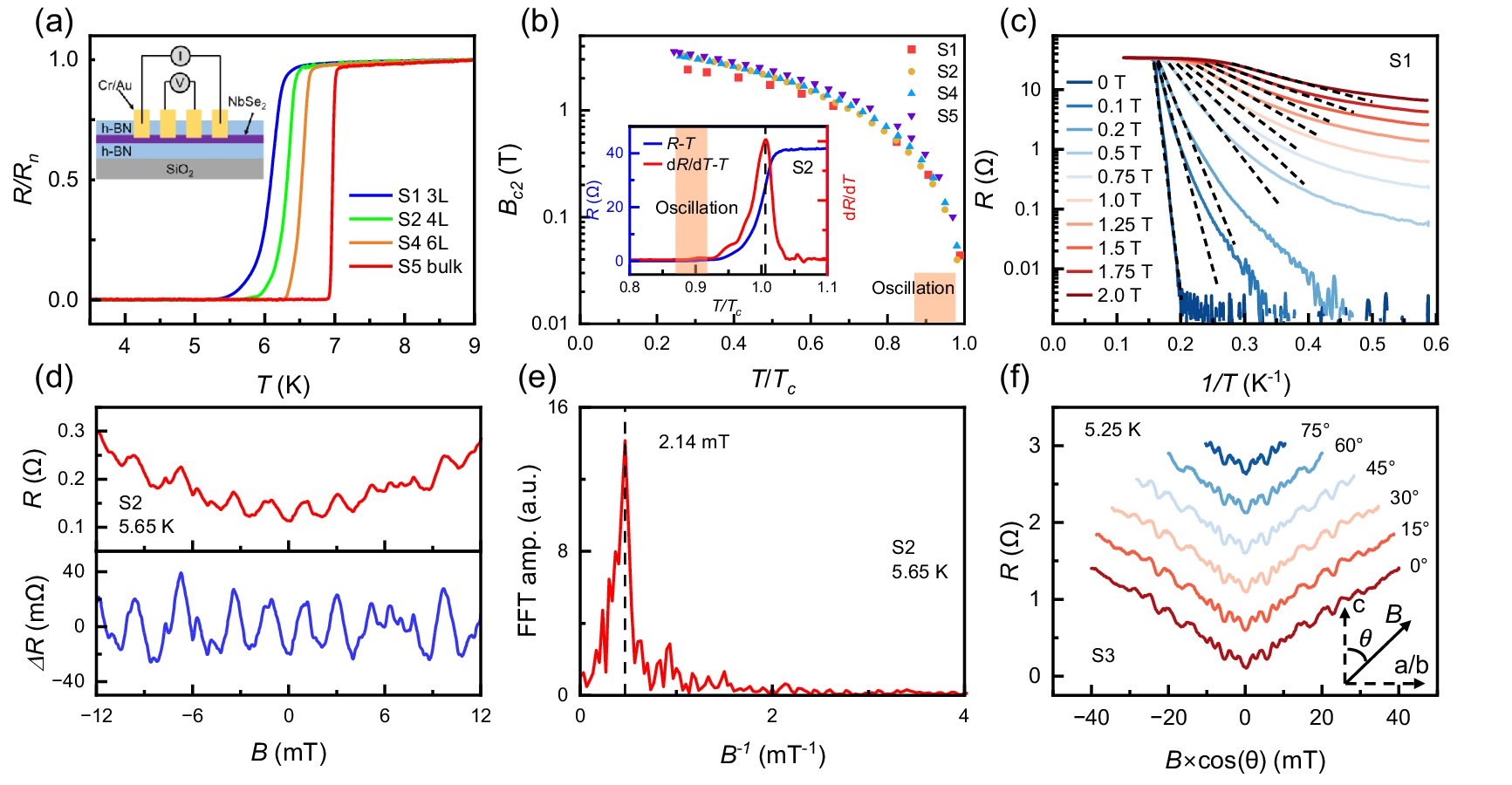}
    \caption{\label{fig:epsart1}(a) Temperature dependence of the normalized resistance of NbSe\textsubscript{2} devices of varying thicknesses. The inset shows a schematic of a typical few-layer NbSe\textsubscript{2} device. (b) Temperature dependence of the upper critical field for S1, S2, S4 and S5. The orange shaded area marks the region where magnetoresistance oscillations are observed. Inset shows the normalized temperature dependence of $R$ and $d R / d T$ for S2, with the orange shaded region indicating the temperature range where oscillations appear. (c) Temperature-dependent resistance of S1 under different perpendicular magnetic fields shown as Arrhenius plot. (d) Resistance oscillations as a function of perpendicular magnetic field (top panel) and the corresponding background-subtracted resistance $\Delta{R}$ (bottom panel) for S2. (e) Fast Fourier transform of the $\Delta{R}$ versus B in (d). (f) Resistance oscillations as a function of perpendicular magnetic field component $B \cdot \cos (\theta)$ for S3, where $\theta$ is the angle between the magnetic field and the direction perpendicular to the sample plane. }
\end{figure*}

In this work, we report periodic magnetoresistance oscillations, superconducting interference patterns, and the superconducting diode effect in unpatterned few-layer NbSe\textsubscript{2}. These oscillations differ from previously reported interference-like phenomena in two key aspects: they emerge exclusively in thin samples and only within the superconducting fluctuation regime. Such behavior suggests that the underlying mechanism is fundamentally different from conventional quantum interference effects observed in patterned samples or in regimes with global phase coherence. We attribute the oscillations to superconducting phase fluctuations, which can give rise to interference-like responses even in the absence of global phase rigidity. By analyzing the temperature dependence of the oscillation amplitude and superconducting diode efficiency, we propose a possible physical scenario in which intrinsic supercurrent loops with asymmetric barriers, periodically modulated by the magnetic field, lead to the observed phenomena when activated vortices cross the loops.

The inset of Fig.\hyperref[fig:epsart1]{~\ref*{fig:epsart1}(a)} displays a schematic of the transport device. Few-layer NbSe\textsubscript{2} was mechanically exfoliated onto SiO\textsubscript{2}/Si substrates, followed by the fabrication of an h-BN/NbSe\textsubscript{2}/h-BN device via standard dry transfer and edge contact techniques\cite{doi:10.1126/science.1244358}. Fig.\hyperref[fig:epsart1]{~\ref*{fig:epsart1}(a)} shows the temperature dependence of the normalized resistance ($R$/$R$(9 K)) for samples of varying thicknesses. The superconducting transition temperature $T_{c}$, defined as the point where $R=0.5R_{n}$, decreases with reduced layer number, consistent with pervious reports\cite{Khestanova2018,Xi2016}. Fig.\hyperref[fig:epsart1]{~\ref*{fig:epsart1}(b)} shows the out-of-plane upper critical field $B_{c 2}$ as a function of temperature, from which the coherence length of the samples can be estimated. For devices S1-S4, corresponding to 3-6 layer samples, the coherence length is approximately 8-10 nm, exceeding the sample thickness and indicating that the samples are in the two-dimensional limit (Supplemental Material, Note 2 and Fig. S2\cite{SM}). The BKT temperature $T_{BKT}$ can be determined from the current-voltage ($I-V$) characteristics, where the power-law dependence $V \sim I^\alpha$ yields $\alpha=3$ (Supplemental Material, Fig. S3\cite{SM}). The extracted $T_{BKT}$ values lie slightly below $T_c$. Moreover, the normal state resistance just above the superconducting transition is much smaller than $h /(2 e)^2$, indicating the high quality and low disorder of the samples\cite{PhysRevLett.42.1165}.

Atomically thin NbSe\textsubscript{2} is known to host an anomalous metal state when its thickness is reduced to the two-dimensional limit\cite{PhysRevLett.134.066002,doi:10.1126/sciadv.aau3826,Benyamini2019,Banerjee2019,Tsen2016}. It is manifested as a fragile, nearly dissipationless superconducting state typically attributed to superconducting fluctuations. Fig.\hyperref[fig:epsart1]{~\ref*{fig:epsart1}(c)} displays the temperature-dependent resistance of S1 in an Arrhenius plot. Under an applied magnetic field, the normal to superconducting transition initially follows a linear, thermally activated behavior. As the temperature decreases, the deviation from linearity and resistance saturation suggest the emergence of an anomalous metal phase. The broader superconducting transition as well as increasingly pronounced anomalous metal phase in few-layer NbSe\textsubscript{2} suggest enhanced superconducting fluctuations in thin layers relative to the bulk (Supplemental Material, Note 4 and Fig. S4\cite{SM}).

Our main observation is that in the superconducting fluctuation regime, pronounced magnetoresistance oscillations emerge within a small magnetic field range, as depicted in  Fig.\hyperref[fig:epsart1]{~\ref*{fig:epsart1}(d)}. The bottom panel shows the background-subtracted oscillations, which are independent of the choice of background subtraction method (Supplemental Material, Note 3\cite{SM}). Magnetoresistance oscillations are observed only within a very narrow temperature range, where the sample is close to entering the zero-resistance superconducting state, as shown in the inset of Fig.\hyperref[fig:epsart1]{~\ref*{fig:epsart1}(b)}.
In Fig.\hyperref[fig:epsart1]{~\ref*{fig:epsart1}(e)}, a fast Fourier transformation (FFT) of the magnetoresistance curves reveals a pronounced peak at 0.466 mT\textsuperscript{-1}, corresponding to an oscillation period of 2.14 mT and an area of 0.97 um\textsuperscript{2}, which is less than the area in between the voltage leads. Fig.\hyperref[fig:epsart1]{~\ref*{fig:epsart1}(f)} shows the angular dependence of the magnetoresistance, plotted against the perpendicular magnetic field component $B \cdot \cos (\theta)$, where $\theta$ is the angle between the magnetic field and the direction perpendicular to the sample plane. The magnetoresistance curves at different angles almost overlap with one another, indicating that the oscillations are primarily due to the perpendicular magnetic field.

\begin{figure}[b]
    \centering
    \includegraphics[width=\columnwidth]{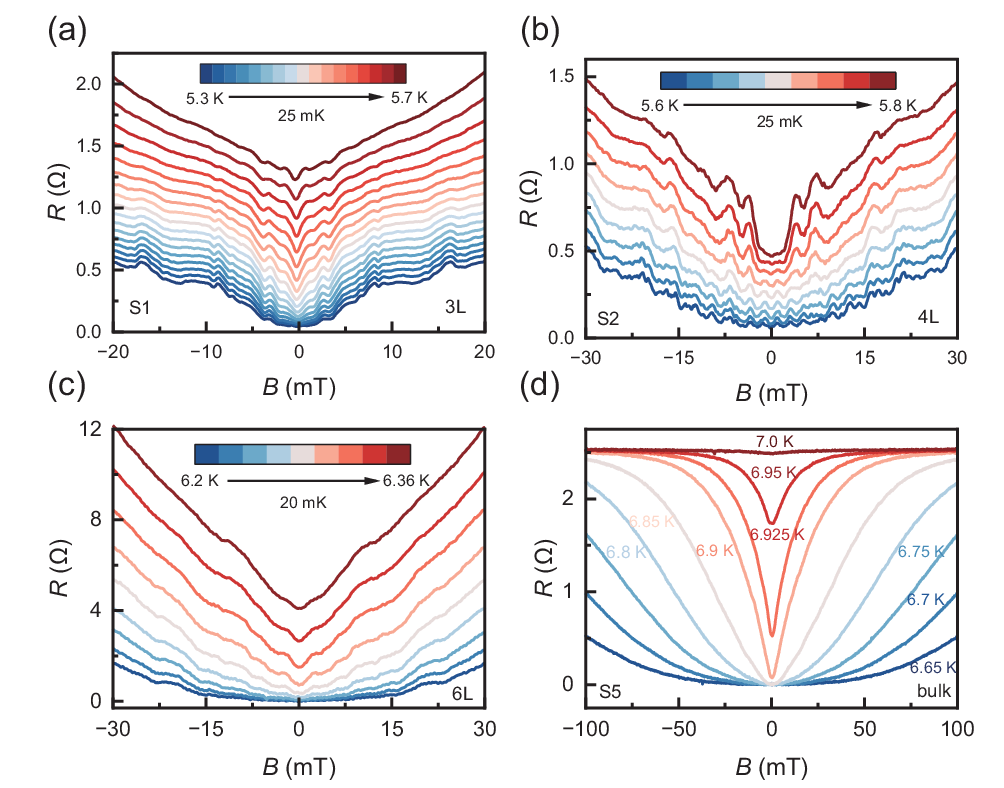} 
    \caption{\label{fig:epsart2}Thickness dependence of the magnetoresistance oscillation. (a-d) Resistance oscillations as a function of perpendicular magnetic field for a trilayer (a), four-layer (b), six-layer (c) and bulk (d) NbSe\textsubscript{2} device at various temperatures. }
\end{figure}

\begin{figure*}[t]
    \centering
    \includegraphics[width=0.8\textwidth]{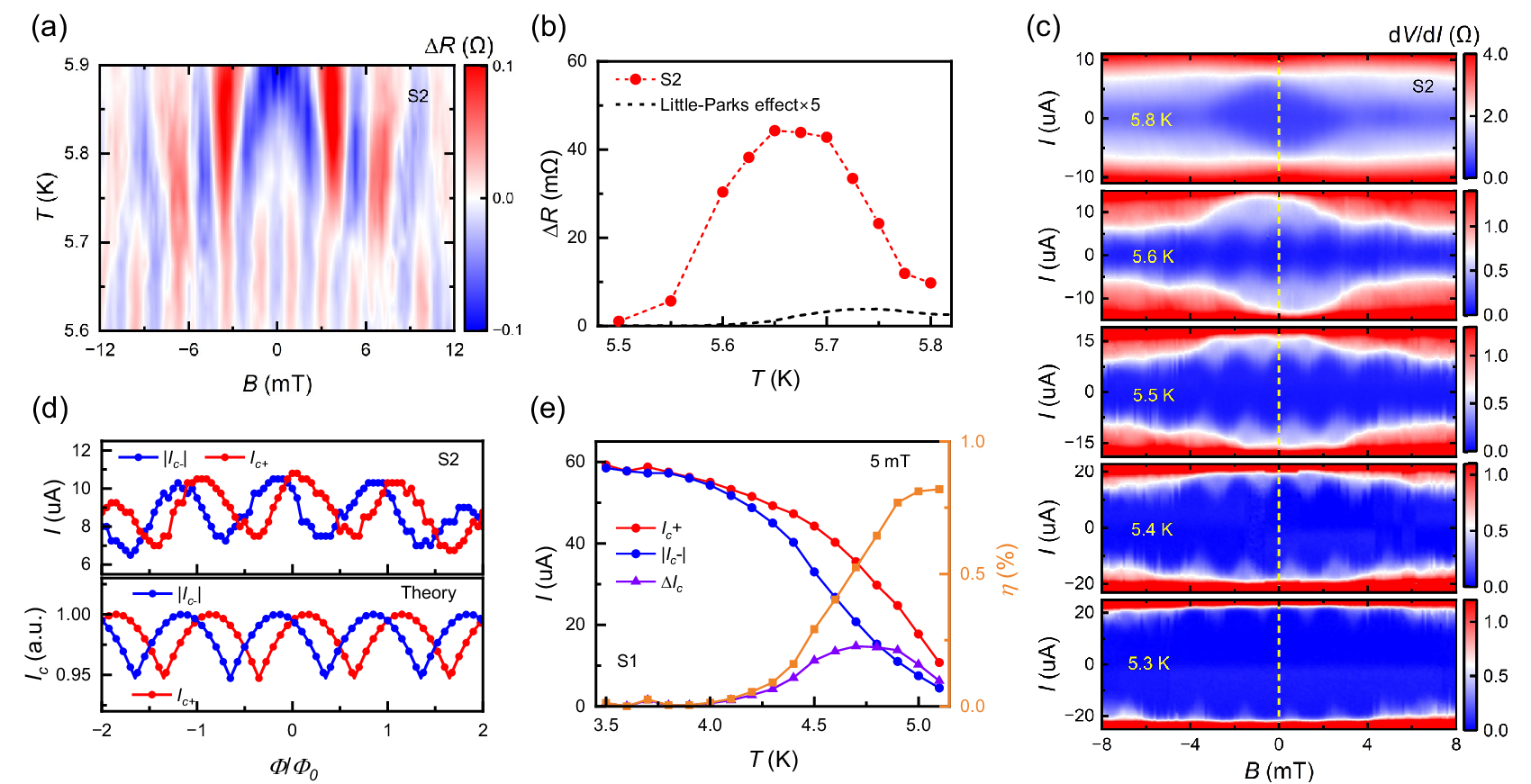}
    \caption{\label{fig:epsart3}(a) Colour maps of the background-subtracted resistance for S2 (4L) as a function of temperature and perpendicular magnetic field. (b) Oscillation amplitude as a function of temperature for device S2. The dashed curve represents five times the amplitude expected from the Little–Parks effect. (c) Colour maps of the differential resistance ($dV/dI$) of S2 as a function of magnetic field and DC bias current at different temperatures. (d) The upper panel shows the field dependence of $I_c^{+}$ and $\left|I_c^{-}\right|$, where the critical currents are defined at $\sim 0.5 \%$ of the normal resistance. The lower panel presents the calculated flux-dependent critical currents $I_c^{+}$ and $\left|I_c^{-}\right|$. (e) Temperature dependence of $I_c^{+}$, $\left|I_c^{-}\right|$, $\Delta I_c$ and $\eta$ for S1 at magnetic field $B=5 \mathrm{mT}$.}
\end{figure*}

We next investigate the thickness dependence of magnetoresistance oscillations in NbSe\textsubscript{2}, as shown in Fig.\hyperref[fig:epsart2]{~\ref*{fig:epsart2}(a-d)}. Notably, the measurements reveal that the periodic oscillations are absent in bulk samples (S5) but become pronounced in thin ones (S1-S4). Furthermore, as the thickness increases, the temperature window over which the oscillations can be observed progressively narrows (Supplemental Material, Fig. S7\cite{SM}). This trend is consistent with the enhanced superconducting fluctuations in thinner layers.

The emergence of a voltage drop in superconductors can originate from the temporal evolution of the phase difference between the two voltage leads, described by: $V=(\hbar / 2 e)(d \Delta \phi / d t)$. Although this phase–voltage relation was originally derived and experimentally demonstrated in Josephson junctions, it is a universal consequence of superconducting phase dynamics\cite{tinkham2004introduction,josephson1962possible}. As such, it applies broadly to any dissipative process involving time-dependent phase evolution. Depending on whether global phase coherence is preserved, two distinct mechanisms can give rise to magnetoresistance oscillations near the superconducting transition. One corresponds to the conventional Little-Parks effect, in which global phase coherence remains intact. In this case, the oscillations originate from the periodic modulation of the superconducting Gibbs free energy with magnetic flux. The second category involves situations where global phase coherence is destroyed. Examples include phase slip centers in 1D superconducting wires\cite{skocpol1974phase,PhysRevB.5.864,PhysRevLett.20.461}, phase slip lines arising from kinematic vortex crossings in 2D superconducting films\cite{PhysRevLett.91.267001,paradiso2019phase}, and other vortex dynamics driven processes such as flux flow and vortex creep. Such phase defects are particularly likely to arise in systems with strong phase fluctuations, such as thin layers of NbSe\textsubscript{2}. In this regime, the reduced superfluid stiffness significantly lowers the energy barriers for phase jumps and for the creation or unbinding of free vortices. 

To elucidate the origin of the magnetoresistance oscillations, we further investigated their temperature dependence, as shown in Fig.\hyperref[fig:epsart3]{~\ref*{fig:epsart3}(a)}. The oscillations persist over a temperature range of $\sim 0.3~\mathrm{K}$, with their period varying at different temperatures, as evidenced by the temperature dependent Fast Fourier transform(Supplemental Material, Fig. S8\cite{SM}). As the temperature is further reduced, the oscillations are progressively suppressed and eventually vanish. One possible mechanism underlying the observed periodic oscillations is the Little–Parks effect. However, conventional Little-Parks effect gives rise to a single oscillation period that is only determined by the dimension of the mesoscopic ring and remains independent of temperature and magnetic field. This is inconsistent with our observations that as the temperature increases, the FFT peak systematically shifts toward lower frequencies. In addition, the amplitude of magnetoresistance oscillations due to conventional Little-Parks effect can be expressed as\cite{PhysRevLett.9.9,Liao2022,PhysRevB.41.2593,gandit1988magnetic}: $\Delta R = (T_c \pi^2 \xi_0^2 / 16 a^2)\,(dR/dT)$, where $a$ denotes the side length of superconducting square loop and can be estimated from the oscillation period via $\Delta B = \Phi_0 / a^2$, with $\Phi_0$ being the flux quantum. As shown in Fig.\hyperref[fig:epsart3]{~\ref*{fig:epsart3}(b)}, the calculated temperature dependence of the oscillation amplitude expected from the Little–Parks effect is substantially smaller than the experimental values, indicating a different mechanism should be responsible for the oscillations. Moreover, even if a superposition of multiple conventional Little–Parks oscillations is considered, the amplitude of each oscillation is expected to peak at the temperature where $\mathrm{d} R / \mathrm{d} T$ is maximal. In our experiment, however, the temperature at which $\mathrm{d} R / \mathrm{d} T$ reaches its maximum is significantly higher than the temperature range over which the oscillations are observed. This mismatch makes it unlikely that the observed magnetoresistance oscillations arise from a superposition of conventional Little–Parks oscillations.

We further measured the differential resistance $\mathrm{d} V / \mathrm{d} I$ as a function of magnetic field $B$ and DC current at various temperatures, revealing the emergence of superconducting interference patterns, as shown in Fig.\hyperref[fig:epsart3]{~\ref*{fig:epsart3}(c)}. Within these interference patterns, we identify a magnetic-field-modulated superconducting diode effect. Similar nonreciprocal transport behavior has been observed in WS\textsubscript{2} nanotubes\cite{qin2017superconductivity}, Josephson junctions\cite{baumgartner2022supercurrent,li2024interfering}, and nano-SQUIDs\cite{doi:10.1126/sciadv.adw4898}, which we hereafter refer to as the interfering diode. The upper panel of Fig.\hyperref[fig:epsart3]{~\ref*{fig:epsart3}(d)} displays the extracted critical currents, $I_c^{+}$ and $\left|I_c^{-}\right|$ from the differential resistance map of S2 at 5.5 K, showing that they do not overlap and that their relative magnitudes are periodically modulated by the external field. The temperature dependence of interfering diode, as shown in Fig.\hyperref[fig:epsart3]{~\ref*{fig:epsart3}(e)}, similarly emerges only within a narrow temperature window corresponding to the superconducting fluctuation regime. This temperature dependence, unlike the monotonic behavior typically observed in the previous reported NbSe\textsubscript{2} nanobridges\cite{bauriedl2022supercurrent} and most other systems, becomes negligible at low temperatures. Such behavior suggests a common origin with the periodic oscillation.
\begin{figure}[b]
    \centering
    \includegraphics[width=0.8\columnwidth]{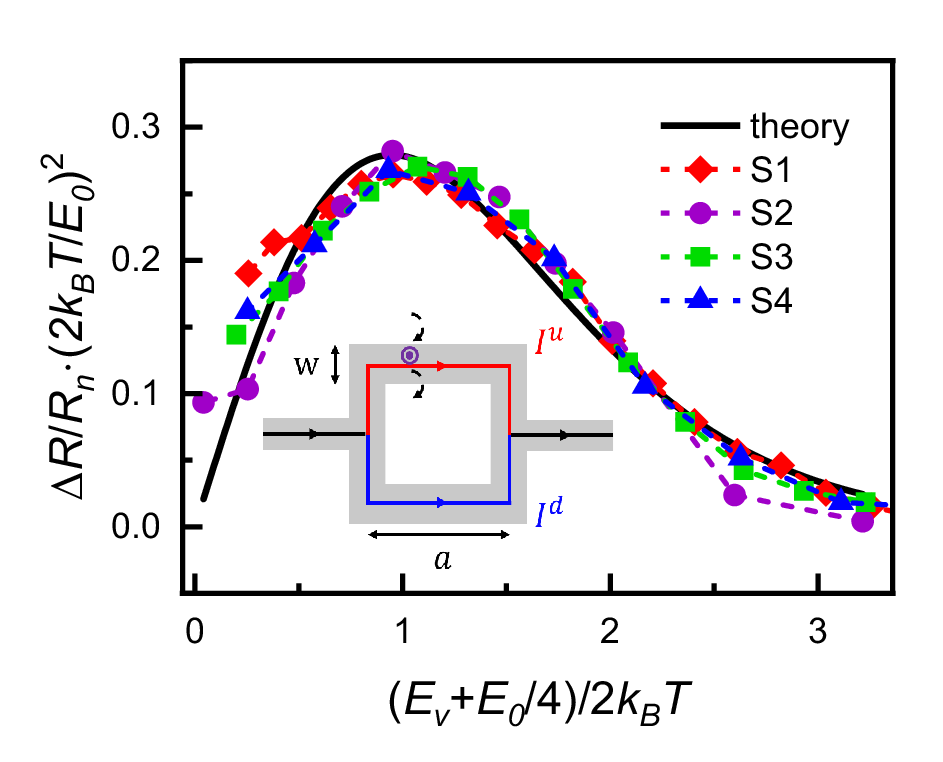} 
    \caption{\label{fig:epsart4}Non-monotonic temperature dependence of the oscillation amplitude for four devices (S1, S2, S3, S4), plotted as $\Delta R / R_n \cdot\left(2 k_B T / E_0\right)^2$ versus $(E_v + E_0/4)/(2k_B T)$. The experimental data for each device are distinguished by different symbols, while the solid black line represents the theoretical model.}
\end{figure}

Vortices are known to exist in few-layer NbSe\textsubscript{2} and are more easily thermally activated than in the bulk\cite{Benyamini2019,Zhang2020,1fzm-pb1d}. This motivates a vortex-based interpretation of the oscillations. We thus attribute the unusually large oscillation amplitude to thermally activated vortices moving in and out of effective supercurrent loops formed by the inhomogeneous current distribution between the voltage leads, as schematically drawn in the inset of Fig.\hyperref[fig:epsart4]{~\ref*{fig:epsart4}}. The activation energy for vortex entry and exit oscillates periodically with the perpendicular magnetic field due to the interference between screening currents and vortex-induced phase winding\cite{PhysRevB.82.094513,Sochnikov2010,PhysRevB.105.224510,PhysRevB.87.081104}. This periodic modulation of the activation barrier naturally leads to a non-monotonic temperature dependence of magnetoresistance oscillations whose amplitude can be expressed as\cite{Liao2022,Sochnikov2010} (Supplemental Material, Note 7\cite{SM}):
\begin{equation}
\Delta R = R_n \left( \frac{E_0}{2k_B T} \right)^{2}
\frac{I_1\!\left[\,(E_v + E_0/4)/(2k_B T)\,\right]}
     {\left[I_0\!\left[\,(E_v + E_0/4)/(2k_B T)\,\right]\right]^{3}},
\end{equation}

where $R_n$ is the nominal resistance that contributes to the oscillation, and $I_0$ and $I_1$ are the zero- and first-order modified Bessel function of the first kind, respectively. The first energy quantity, $E_v = \Phi_0^2/(2 \pi \mu_0 \Lambda(T)) \cdot \ln(2w/(\pi \xi(T)))$, represents the energy required for vortex creation and is independent of the external field. The second energy quantity, $E_0 = \Phi_0^2/(2 \pi \mu_0 \Lambda(T)) \cdot w/a$, characterizes the interaction of a vortex with the current associated with the fluxoid. Here, $w$ is the annulus width of square loops, $\mu_0$ is the vacuum permeability. $\xi(T) = 0.74\,\xi_0\,(1 - T/T_c)^{-1/2}$ is the Ginzburg-Landau coherence length. $\Lambda(T) = 2 \, \lambda(T)^2 / d$ is the Pearl penetration depth, where $d$ is the sample thickness and $\lambda(T) = \lambda_0 \, (1 - (T/T_c)^2)^{-1/2}$ is the London penetration depth. As can be seen from the Eq (1), the superfluid stiffness $\rho_s$ becomes small and global phase coherence is lost near $T_c$ $\left(\rho_s \propto E_v\right)$, leading to a strongly suppressed amplitude. At low temperatures, by contrast, the energy barrier for thermally activated vortex crossings becomes too large, effectively freezing out vortex motion and again reducing the amplitude. This leaves the oscillation only observable at superconducting fluctuation regime. By fitting the temperature dependence of oscillation, we obtained $w$, $R_n$ and $\Lambda_0 = 2 \, \lambda_0^2 / d$, while $\xi_0$, $T_c$ and $a$ were treated as fitting parameters. The oscillation amplitudes of all four devices are well captured by the model, as shown in Fig.\hyperref[fig:epsart4]{~\ref*{fig:epsart4}}. The extracted Pearl penetration depths for each of the device, summarized in Table S1 of the Supplemental Material\cite{SM}, decrease systematically with increasing sample thickness, consistent with reported in scanning SQUID measurements\cite{Fridman2025}.

The traversing vortices can also naturally account for the observed interference patterns. 
Driven by the Lorentz force, kinematic vortices traversing the superconductor act as dynamic phase-slip lines, effectively serving as weak link analogous to that in Josephson junction\cite{PhysRevLett.91.267001,PhysRevResearch.2.043204}. Such vortex-mediated phase-slip processes may thus enable superconducting interference phenomena in the strong phase-fluctuation regime of few-layer NbSe\textsubscript{2}. As the temperature changes, both vortex density and supercurrent paths are modified, changing the configuration and dynamics of vortices. This evolution can drive the interference pattern to evolve from a double-slit–like pattern into a Fraunhofer-like pattern. At sufficiently low temperatures, thermally activated vortex motion is suppressed, and the interference pattern correspondingly disappears (Supplemental Material, Fig. S9\cite{SM}).

To account for the superconducting diode effect, we assume supercurrent loop is asymmetric, thus giving rise to an asymmetric potential. In such an asymmetric loop, the supercurrents flowing through the upper and lower segments are unequal, $I^u \neq I^d$, as shown in the inset of Fig.\hyperref[fig:epsart4]{~\ref*{fig:epsart4}}. This will generate an additional magnetic flux due to self-inductance, which leads to a flux phase shift that modifies the energy of the supercurrent loop for opposite current directions\cite{Le2024}. Following previous analyses\cite{Sochnikov2010,PhysRevB.82.094513,PhysRevB.69.064516,PhysRevB.68.214505}, we develop a minimal model based on an asymmetric supercurrent loop. In this framework, the effective energy barrier for thermally activated vortices crossing the supercurrent loop depends on current direction and can be expressed as:
\begin{equation}
\Delta E^{ \pm}=E_v+E_0\left[ \pm\left(N-H / H_0+\kappa \operatorname{sign}(I)\right)+1 / 4\right],
\end{equation}
where $E_v$ and $E_0$ have been discussed above. $N$ denotes the winding number of the supercurrent loop. The coefficient $\kappa$ characterizes the magnitude of the phase shift, and $\operatorname{sign}(I)$ represents the sign function of the bias current $I$. Based on the vortex dynamics model, we numerically calculate the flux-dependent critical currents  $I_c^{+}$ and $\left|I_c^{-}\right|$, as shown in the lower panel of Fig.\hyperref[fig:epsart3]{~\ref*{fig:epsart3}(d)} (Supplemental Material, Note 7\cite{SM}). At zero bias current, the model reproduced magnetoresistance oscillations without a phase shift, whose oscillation amplitude follows Eq. (1).  

In summary, we have observed periodic magnetoresistance oscillations in few-layer NbSe\textsubscript{2} within the superconducting fluctuation regime, where global phase coherence is destroyed. The origin of these oscillations is distinct from the conventional Little–Parks effect, which requires preserved global phase coherence. Our results reveal that superconducting interference and superconducting fluctuations, phenomena traditionally associated with the presence or absence of global phase coherence, can coexist. Our findings strongly suggest that enhanced fluctuations, which make the superconducting state fragile in the two-dimensional limit, play a central role in enabling emergent quantum interference phenomena.

\begin{acknowledgments}
\vspace{10pt}
Acknowledgments---We thank Kun Jiang, Menghan Liao and Noah F. Q. Yuan for invaluable discussions. This work was supported by the National Natural Science Foundation of China (Grants No. 12404549, No. 12574523 and No. 12474143), the National Key Research and Development Program of China (Grant No. 2021YFA1400400), the Guangdong Fundamental Research Program (Grants No. 2025A1515012039), the Shenzhen Fundamental Research Program (Grants No. JCYJ20220818100405013 and No. JCYJ20230807093204010), and Guangdong Provincial Quantum Science Strategic Initiative (GDZX2401010).
\end{acknowledgments}
\vspace{10pt}
Data availability---The data that support the findings of this article are not publicly available. The data are available from the authors upon reasonable request.


\bibliographystyle{apsrev4-2}

\bibliography{apssamp}

\end{document}